\documentclass{article}
\usepackage{graphicx} 

\usepackage{amsmath, amsfonts, amssymb, amsthm, mathrsfs, physics, braket}
\usepackage{comment}

\usepackage{hyperref} 
\hypersetup{colorlinks=true, citecolor=blue, linkcolor=blue, urlcolor=blue} 

\usepackage[title]{appendix} 

\newcommand{\domain}{\mathcal{D}}
\newcommand{\operator}[1]{\mathsf{#1}}
\newcommand{\hilbertspace}{\mathcal{H}}

\title{Characteristic time operators as quantum clocks}
\author{Ralph Adrian E. Farrales* and Eric A. Galapon \\ 
\\
Theoretical Physics Group, National Institute of Physics \\ University of the Philippines Diliman, Philippines \\ \\
*Corresponding author: refarrales@up.edu.ph}
\date{\today}

\begin{document}

\maketitle

\begin{abstract}
We consider the characteristic time operator $\operator{T}$ introduced in [E. A. Galapon, \textit{Proc. R. Soc. Lond. A}, \textbf{458}:2671 (2002)] which is bounded and self-adjoint. For a semibounded discrete Hamiltonian $\operator{H}$ with some growth condition, $\operator{T}$ satisfies the canonical relation $[\operator{T},\operator{H}]\ket{\psi}=i\hbar\ket{\psi}$ for $\ket{\psi}$ in a dense subspace of the Hilbert space. While $\operator{T}$ is not covariant, we show that it still satisfies the canonical relation in a set of times of total measure zero called the time invariant set $\mathscr{T}$. In the neighborhood of each time $t$ in $\mathscr{T}$, $\operator{T}$ is still canonically conjugate to $\operator{H}$ and its expectation value gives the parametric time. Its two-dimensional projection saturates the time-energy uncertainty relation in the neighborhood of $\mathscr{T}$, and is proportional to the Pauli matrix $\sigma_y$. Thus, one can construct a quantum clock that tells the time in the neighborhood of $\mathscr{T}$ by measuring a compatible observable.
\\
\\
\noindent
keywords: quantum clock, characteristic time operator, time-energy canonical commutation relation
\end{abstract}

\section{Introduction}
A time operator $\operator{T}$ is, by definition, canonically conjugate to the Hamiltonian $\operator{H}$, meaning that this pair of operators satisfy the time-energy canonical commutation relation (TECCR)
\begin{equation} \label{eq:teccr}
    [\operator{T},\operator{H}] = i\hbar \, .
\end{equation}
From the Heisenberg equation of motion $\dv*{\operator{A}}{t}= (i\hbar)^{-1} [\operator{A},\operator{H}]$, it was then typically understood that time operators move in step with parametric time $t$ at all moments, implying that $\operator{T}$ satisfies the covariance property $\operator{T}(t) = \operator{T}(0) + t$ for all times. The study of such an operator $\operator{T}$ has had quite a dramatic history \cite{muga2008time,muga2009time}. It was initially thought that observables should be represented by self-adjoint operators in quantum mechanics. Very early on however, Pauli argued that for a self-adjoint operator $\operator{T}$ to satisfy \eqref{eq:teccr}, the spectrum of the Hamiltonian must span the entire real line. This called for the historical rejection of the existence of any such $\operator{T}$ given the semibounded Hamiltonians of quantum mechanics \cite{pauli1980general,holevo1978estimation,srinivas1981time,busch1995operational,giannitrapani1997positive,toller1999localization}. With the later discovery of positive-operator-valued measures (POVM), it was shown that quantum observables need not be self-adjoint, and allowed time to be accommodated back into quantum mechanics as a non-self-adjoint, covariant, POVM observable \cite{srinivas1981time, busch1995operational, giannitrapani1997positive, toller1999localization}. Throughout the years, much progress has been made on time POVMs, but recently, there has been renewed interest on the study of self-adjoint time operators. There arose multiple counterexamples to Pauli's claim \cite{nelson1959analytic,kraus1965remark,garrison1970canonically,reed1975fourier,galindo1984phase}, but it all culminated into a work by one of us showing that indeed, there actually is no inconsistency in assuming a self-adjoint operator canonically conjugate to a semibounded or discrete Hamiltonian \cite{galapon2002pauli}. This provided a new direction in tackling the problem of time in quantum mechanics, which essentially tells us to reconsider the previously neglected self-adjoint time operator solutions to the TECCR.

The crux of the problem arises by first assuming that the TECCR holds in the entire Hilbert space $\hilbertspace$, and second by assuming that the TECCR holds for all times. Firstly, the TECCR actually only holds in a proper subspace $\domain_c$ of $\hilbertspace$ called the canonical domain \cite{galapon2002pauli,galapon2006could,galapon2009post}. Solutions to the TECCR are then the triple $\mathcal{C}(\operator{T},\operator{H}, \domain_c)$ where the operators $\operator{T}$ and $\operator{H}$ satisfy
\begin{equation} \label{eq:teccrdc}
    [\operator{T},\operator{H}]\ket{\psi} = i\hbar \ket{\psi} \, ,
\end{equation}
for all $\ket{\psi} \in \domain_c$. In the same Hilbert space, there could exist multiple solutions to the TECCR, wherein different classes of solutions each satisfy different properties. The solutions can be classified into two: the dense-category solutions when $\domain_c$ is dense, and the closed-category solutions when $\domain_c$ is closed. Within each category is a further split into different classes of canonical pairs (and their unitary equivalents). This fact is in line with the multifaceted nature of time, wherein different classes of time observables each satisfy different properties \cite{srinivas1981time,galapon2006could,busch2008time,arai2020inequivalent}. 

Pauli's conditions (that both operators form a system of imprimitivities in the real line \cite{giannitrapani1997positive,toller1999localization,galapon2002pauli,mackey1963infinite}) only hold for specific systems wherein both operators are unbounded and satisfy the Weyl relation, which cannot be forced as a requirement for bounded operators \cite{kraus1965remark,garrison1970canonically,galapon2006could}. As another example, the construction of a quantum clock does not require imprimitivity. According to Garrison and Wong \cite{garrison1970canonically}, a quantum clock satisfies the following properties: there should exist (i) a family of states $\ket{\Lambda(\tau)}$ satisfying $\operator{U}_t \ket{\Lambda(\tau)} = \ket{\Lambda(\tau+t)}$ for $\operator{U}_t = \exp(-it\operator{H}/\hbar)$, and (ii) a self-adjoint operator $\operator{T}$ whose measurement on $\ket{\Lambda(\tau)}$ yields $\tau$ with negligible dispersion. In \cite{garrison1970canonically}, the harmonic oscillator phase and number operators were considered, which can be mapped to a time operator and Hamiltonian as a canonical pair in its canonical domain. Throughout the paper, we shall call states $\ket{\Lambda(\tau)}$ satisfying the above properties as \textit{clock states}. As a final example, if covariance is a required property of the system, then the time operators can be treated as POVMs; instead of imprimitivity, the conjugate pair form a system of covariance \cite{giannitrapani1997positive,toller1999localization,castrigiano1980systems}. In contrast, the self-adjoint time operators are covariant only for a completely continuous energy spectrum \cite{arai2008time}, contributing to the pessimism for semibounded or discrete Hamiltonians \cite{galapon2002pauli,galapon2009post}. All these approaches to time in quantum mechanics are just a few of the possible classes, and not the only possible classes, of solutions to the TECCR \cite{galapon2006could,farrales2022conjugates}.

We then arrive at the heart of this work, which addresses the second problem with the TECCR, in that it is assumed that all time operators satisfy \eqref{eq:teccrdc} for all times. In this view, one would find covariant time operators such as time POVMs as the only acceptable type of time operators in quantum mechanics, since self-adjoint time operators cannot be canonically conjugate to a semibounded discrete Hamiltonian for all times \cite{arai2008time,hall2008almost}. Any attempt in constructing such a non-covariant operator is questioned on its physical interpretation \cite{dodonov2015energy,sainadh2020attoclock}. In this letter, we thus look into the self-adjoint time operators of \cite{galapon2002self}: to every discrete semibounded Hamiltonian with some growth condition, there exists conjugate to it a \textit{characteristic} self-adjoint operator, whose construction only depends on the energy eigenkets and eigenvalues. This non-covariant operator has been the subject of much study \cite{arai2020inequivalent,arai2008time,arai2005generalized,arai2009necessary}, wherein we highlight \cite{galapon2006could,caballar2009characterizing} as it was suggested that the characteristic time operator behaves as a quantum clock. In particular, the work of \cite{caballar2009characterizing} showed that for a particle in a box, the time-evolved eigenstates of the characteristic time operator have a high probability of transitioning to another eigenstate at an instant of time equal to the difference of their corresponding eigenvalues, assuming this difference is sufficiently small. A quantum clock can be any quantum system which evolves from one state to another, wherein measurement of some dynamical observable (a clock pointer) with a known time dependence helps us infer the time that has elapsed \cite{garrison1970canonically,galindo1984phase,busch2008time,wigner1957relativistic,salecker1958quantum,susskind1964quantum,peres1980measurement,mayato2008quantum,ludlow2015optical}. One thing to note though is that the clock pointer observable is not the time operator itself, and thus, a theory on quantum clocks using self-adjoint time operators is still lacking. 

Thus, in this letter, we look at the characteristic time operator and show that it indeed acts as a quantum clock, giving us access to the parametric time only in the neighborhood of a set of times of total measure zero. We first give a brief review in Section \ref{sec:dc} on the characteristic time operator as a TECCR solution and its corresponding canonical domain. We then study in Section \ref{sec:teccrsoln} some properties of this non-covariant operator. Firstly, we show in Section \ref{sec:tis} that it satisfies the TECCR in a set of times of total measure zero called the time invariant set. Additionally, the TECCR continues to hold in the neighborhood of the time invariant set, as seen in Section \ref{sec:local}. For times near the time invariant set, we shall see in Section \ref{sec:evvar} that the expectation value of the characteristic time operator gives the parametric time, while its two-dimensional projection saturates the time-energy uncertainty relation. The characteristic time operator then behaves as a quantum clock for times near the time invariant set, and we provide the Larmor clock case in Section \ref{sec:qclock} as an example. Finally, we shall give a discussion of our results in Section \ref{sec:disc}.

\section{Characteristic time operator and its canonical domain} \label{sec:dc}
Let $\operator{H}$ be a non-degenerate Hamiltonian with eigenkets $\ket{s}$ and eigenvalues $E_s$ for $s = 0, 1, 2, \dotsc$, explicitly written as
\begin{equation}
    \operator{H} = \sum_{s=0}^\infty E_s \ket{s} \bra{s} \, .
\end{equation}
Let the Hilbert space be spanned by the eigenkets $\ket{s}$,
\begin{equation}
    \hilbertspace = \qty{ \ket{\varphi} = \sum_{s=0}^\infty \varphi_s \ket{s} , \sum_{s=0}^\infty \abs{\varphi_s}^2 < \infty } \, ,
\end{equation}
and thus, the domain of $\operator{H}$ is explicitly given by
\begin{equation}
    \domain_{\operator{H}} = \qty{ \ket{\varphi} = \sum_{s=0}^\infty \varphi_s \ket{s} \in \hilbertspace , \sum_{s=0}^\infty E_s^2 \abs{\varphi_s}^2 < \infty } \, .
\end{equation}
One can then construct the characteristic time operator for the non-degenerate Hamiltonian
\begin{equation} \label{eq:cto}
    \operator{T} = \sum_{s \neq s' \geq 0}^\infty \frac{i}{\omega_{s,s'}} \ket{s} \bra{s'} \, ,
\end{equation}
where $\omega_{s,s'} = (E_s - E_{s'})/\hbar$, with the condition that $\sum_{s=0}^\infty E_s^{-2} < \infty$. The domain of $\operator{T}$ is explicitly
\begin{equation}
    \domain_\operator{T} = \qty{ \ket{\varphi} = \sum_{s=0}^N \varphi_s \ket{s}, \varphi_s \in \mathcal{C}, N < \infty } \, .
\end{equation}

We look at the pair $\operator{T}$ and $\operator{H}$ in a subspace $\domain_c$ of $\domain_\operator{TH} \cap \domain_\operator{HT}$. For $\ket{\varphi} \in \domain_c$, $(\operator{T}\operator{H} - \operator{H}\operator{T})\ket{\varphi} = -i\hbar \sum_{s=0}^\infty \sum_{s'\neq s}^N \varphi_{s'} \ket{s}$. The condition $\sum_{s=0}^N \varphi_s = 0$ then gives us the desired relation $(\operator{T}\operator{H} - \operator{H}\operator{T})\ket{\varphi} = i\hbar\ket{\varphi}$. The required condition is satisfied with the following choice: $\varphi_s = \sum_{k=0}^{s-1} a_{s,k} - \sum_{k=s+1}^N a_{k,s}$, where the $a_{s,k}$'s are constants. We then have the subspace of $\hilbertspace$
\begin{equation} \label{eq:ctodc}
    \domain_c = \qty{ \ket{\varphi} = \sum_{l=0}^{N-1} \sum_{k = l+1}^N a_{kl} (\ket{k} - \ket{l}), a_{kl} \in \mathcal{C}, N < \infty} \, ,
\end{equation}
as our canonical domain. This subspace $\domain_c$ is dense since the only vector orthogonal to it is the zero vector. The pair $\operator{T}$ and $\operator{H}$ is then canonically conjugate in this dense canonical domain $\domain_c$. The characteristic time operator is essentially self-adjoint, meaning there exists a unique self-adjoint extension $\operator{\bar{T}} = (\operator{T}^*)^*$ which remain canonically conjugate with $\operator{H}$ in the same dense canonical domain $\domain_c$. Thus, the triple $\mathcal{C}(\operator{\bar{T}},\operator{H}, \domain_c)$ constitute a TECCR solution of dense-category consisting of a self-adjoint canonical pair \cite{galapon2002self}. We will show in the succeeding section that elements of the canonical domain can in fact act as the clock states of our quantum clock.

\section{A TECCR solution in the neighborhood of a set of measure zero} \label{sec:teccrsoln}
\subsection{Time invariant set} \label{sec:tis}
The canonical domain is not invariant under time translation. Consider the vector
\begin{equation}
    \ket{k,l} = \ket{k} - \ket{l} \, ,
\end{equation}
where $k \neq l$, which is in $\domain_c$. We see that
\begin{equation}
    \operator{U}_t \ket{k,l} = \exp(-i\omega_k t) \ket{k} - \exp(- i \omega_l t) \ket{l} \, .
\end{equation}
It is not always true that the coefficients vanish for all times $t$, thus, the time evolution of $\ket{k,l}$ pushes it outside the canonical domain. However, for a given $k$ and $l$, there are times wherein $\operator{U}_t \ket{k,l}$ does go back to $\domain_c$. These times can be obtained by determining the common period of $\exp(-i\omega_k t)$ and $\exp(-i\omega_l t)$,
\begin{equation} \label{eq:tkl}
    \mathscr{T}_{k,l} = \qty{ t : t \omega_{k,l} = 2 n \pi, n = 0, \pm 1, \pm 2, \dotsc } \, .
\end{equation}
We then define
\begin{equation}
    \mathscr{T} = \cap \mathscr{T}_{k,l} \, ,
\end{equation}
as the set of all times common to all $\mathscr{T}_{k,l}$ (for all $k$ and $l$, $k \neq l$), i.e., for every $t$ in $\mathscr{T}$, the evolution $\operator{U}_t \ket{\varphi}$ is in $\domain_c$ for every $ \ket{\varphi}$ in $\domain_c$. We denote $\mathscr{T}$ as the \textit{time invariant set}. Generally, $\mathscr{T}_{kl} \subseteq \mathscr{T}$, and $\mathscr{T}$ at least contains $t = 0$ (since $\ket{\varphi}$ is assumed to be initially at $\domain_c$) and is thus not empty.

Suppose the energy eigenvalues can be written in the form $E_s = \hbar \omega f(s)$ where $f(k) - f(l) = \Delta_{k,l}$ is positive for every $k > l$, and is not necessarily an integer. Let $\vec{\Delta}$ be the set of all $\Delta_{k,l}$'s. In general, we see that the time invariant set is
\begin{equation}
    \mathscr{T} = \qty{t : t = \frac{2j\pi}{\omega \gcd(\vec{\Delta})}, j = 0, \pm 1, \pm 2, \dotsc } \, ,
\end{equation}
where $\gcd$ is the greatest common divisor. 

Let us look at specific cases. Let $\Delta_0 = \min(\vec{\Delta})$ be the smallest $\Delta_{k,l}$ in $\vec{\Delta}$. If all other $\Delta_{k,l}$'s are just multiples of $\Delta_0$, the time invariant set is
\begin{equation}
    \mathscr{T} = \qty{t : t = \frac{2j\pi}{\omega \Delta_0}, j = 0, \pm 1, \pm 2, \dotsc } \, .
\end{equation}
For integer $\Delta_{k,l}$'s, if there exists a $\Delta_{k,l}$ that has no common factors with $\Delta_0$, then
\begin{equation}
    \mathscr{T} = \qty{t : t = \frac{2j\pi}{\omega}, j = 0, \pm 1, \pm 2, \dotsc } \, .
\end{equation}
If there exists a $\Delta_{k,l}$ such that $\Delta_{k,l}$ and $\Delta_0$ are incommensurate (i.e., $\Delta_{k,l}/\Delta_0$ is irrational), then
\begin{equation}
    \mathscr{T} = \qty{ t = 0 } \, .
\end{equation}
Thus, the nature of the energy spectrum, i.e., the nature of the energy differences, determines the time invariant set. 

As an example, consider the harmonic oscillator energy eigenvalues $E_s = \hbar \omega_0 (s + 1/2)$ for $s = 0, 1, 2, \dotsc$. We have $E_s - E_{s'} = \hbar\omega_0 \Delta_{s,s'}$ where $\Delta_{s,s'} = s - s'$, which has the minimum $\Delta_0 = 1$. Since all other $\Delta_{s,s'}$'s are integers and are multiples of $\Delta_0$, then the time invariant set is then the set of all $t$ satisfying $t = 2j\pi/\omega_0$ for integer $j$. Interestingly, we see that it occurs at times that are multiples of the classical period of the harmonic oscillator.

In summary, for $\operator{T}_t = \operator{U}_t^\dagger \operator{T} \operator{U}_t$, the canonical commutation relation $(\operator{T}_t \operator{H} - \operator{H}\operator{T}_t)\ket{\varphi} = i\hbar \ket{\varphi}$ for $\ket{\varphi} \in \domain_c$ holds for every $t$ in the time invariant set $\mathscr{T}$. While the state moves outside $\domain_c$ via time evolution, we shall show  below that the TECCR still holds even in the neighborhood of every $t$ in $\mathscr{T}$.

\subsection{Local dynamics} \label{sec:local}
We next look at the local dynamics of the canonical domain for short times. Let $\domain_\operator{H}$ be the domain of the Hamiltonian $\operator{H}$. Let $\ket{\varphi} \in \domain_\operator{H}$ be written explicitly as
\begin{equation}
    \ket{\varphi} = \sum_s \varphi_s \ket{s} \, ,
\end{equation}
where $\ket{s}$ are the energy eigenkets of the Hamiltonian $\operator{H}$ corresponding to the energy eigenvalue $E_s$. Consider the operator $\operator{I} - i t \operator{H}/\hbar$ defined in $\domain_\operator{H}$. For every $\ket{\varphi} \in \domain_\operator{H}$,
\begin{equation}
\begin{aligned}
    \norm{\operator{U}_t \ket{\varphi} - \qty(\operator{I} - \frac{i}{\hbar} t \operator{H}) \ket{\varphi}}^2 &= 2 \norm{\ket{\varphi}}^2 - 2 \sum_s \abs{ \varphi_s }^2 \cos \omega_s t \\
    &\quad - 2 \sum_s \abs{ \varphi_s }^2 \omega_s t \sin \omega_s t + \frac{1}{\hbar^2} t^2 \norm{ \operator{H} \ket{\varphi} }^2 \, ,
\end{aligned}
\end{equation}
where $\omega_s=E_s/\hbar$. Expanding $\cos\omega_s t$ and $\sin\omega_s t$ shows that
\begin{equation}
    \norm{\operator{U}_t \ket{\varphi} - \qty(\operator{I} - \frac{i}{\hbar} t \operator{H}) \ket{\varphi}}^2 = \order{t^2} \, .
\end{equation}
We then have $\norm{\operator{U}_t \ket{\varphi} - \qty(\operator{I} - i t \operator{H}/\hbar) \ket{\varphi}} = \order{t}$. This means that, up to $\order{t}$, we can approximate $\operator{U}_t$ by $\operator{I} - i t \operator{H}/\hbar$ with the expense of reducing the domain of $\operator{U}_t$ to $\domain_\operator{H}$.

This allows us to write $\operator{T}_t$ in the form $\qty(\operator{I} + i t \operator{H}/\hbar) \operator{T} \qty(\operator{I} - i t \operator{H}/\hbar)$. Originally, $\operator{U}_t^\dagger \operator{T} \operator{U}_t$ is defined in the entire domain of $\operator{T}$ because its domain $\domain_\operator{T}$ is invariant under time translations. But under the replacement $\operator{T}_t = \qty(\operator{I} + i t \operator{H}/\hbar) \operator{T} \qty(\operator{I} - i t \operator{H}/\hbar)$, the domain of $\operator{T}_t$ reduces to $\domain_c$, which is a proper subspace of $\domain_\operator{T}$. Thus, for short times, the operator $\operator{H}\operator{T}_t$ is well defined in the entire $\domain_c$. This allows us to derive the time evolution of $\operator{T}$ for short times
\begin{equation}
    i\hbar \dv{ \ket{\operator{T}_t \varphi} }{t} = \qty[\operator{T}_t,\operator{H}] \ket{\varphi} \, ,
\end{equation}
where $\ket{\varphi} \in \domain_c$. Since $\domain_c$ is stable under short time evolution, we have $\qty[\operator{T}_t,\operator{H}] \ket{\varphi} = i\hbar \ket{\varphi}$, giving
\begin{equation}
    \eval{\dv{ \ket{\operator{T}_t \varphi} }{t}}_{t \in \mathscr{T}} = \ket{\varphi} \, .
\end{equation}
Let $t(\tau) = \tau + t'$ for time interval $\tau$ and $t' \in \mathscr{T}$. Then,
\begin{equation}
    \operator{U}_{t(\tau)}^\dagger \operator{T} \operator{U}_{t(\tau)} \ket{\varphi} = (\operator{T} + \tau + \order{\tau^2}) \ket{\varphi} \, ,
\end{equation}
giving $\operator{U}_{t(\tau)}^\dagger \operator{T} \operator{U}_{t(\tau)} = \operator{T} + \tau$ for short times near the time invariant set $\mathscr{T}$.

In summary, not only does the TECCR hold in a set of times of total measure zero; it also holds true in the neighborhood of each element of that set. Thus, the characteristic time operator $\operator{T}$ evolves in step with parametric time in a set of measure zero. $\operator{T}$ then behaves in the expected way an acceptable time operator should behave around $\mathscr{T}$, but not away from $\mathscr{T}$.

\subsection{Expectation value and variance} \label{sec:evvar}
Let us demonstrate the above results and analyze the quantum dynamics of the characteristic time operators for states $\ket{\varphi_{k,l}} = 2^{-1/2} (\ket{k} - \ket{l})$ for every $k > l$. Note that every $\ket{\varphi_{k,l}}$ belongs to $\domain_c$. For every $t$ in $\mathscr{T}_{kl}$ \eqref{eq:tkl}, the evolved states $\operator{U}_t\ket{\varphi_{k,l}}$ are also in $\domain_c$.

The expectation value and variance of $\operator{T}$ in $\ket{\varphi_{k,l}}$ are
\begin{equation}
    \braket{\varphi_{k,l}|\operator{T}|\varphi_{k,l}} = 0 \, ,
\end{equation}
\begin{equation}
    \Delta T_{k,l}^2 = \frac{1}{\omega_{k,l}^2} + \frac{1}{2} \omega_{k,l}^2 \sum_{s \neq k,l} \frac{1}{\omega_{s,k}^2 \omega_{s,l}^2} \, .
\end{equation}
Let $t(\tau) = \tau + 2n\pi/\omega_{kl}$ for integer $n$, where $2n\pi/\omega_{kl}$ are the times in $\mathscr{T}_{kl}$. The time interval $\tau$ can have values within $[0,2\pi/\omega_{kl})$. For the time-evolved characteristic time operator $\operator{T}_{t(\tau)} = \operator{U}_{t(\tau)}^\dagger \operator{T} \operator{U}_{t(\tau)}$, we have the following expectation value and variance,
\begin{equation}
    \braket{\varphi_{k,l}|\operator{T}_{t(\tau)}| \varphi_{k,l}} = \frac{1}{\omega_{k,l}} \sin \omega_{k,l} \tau \, ,
\end{equation}
\begin{equation}
    \Delta T_{k,l}^2(t(\tau)) = \frac{\cos^2\omega_{k,l} \tau}{\omega_{k,l}^2} + \frac{1}{2} \sum_{s \neq k,l} \qty( \frac{1}{\omega_{s,k}^2} + \frac{1}{\omega_{s,l}^2} - \frac{2 \cos \omega_{k,l} \tau}{\omega_{s,k} \omega_{s,l}} ) \, .
\end{equation}
The values above are periodic in $t$ with period $2\pi /\omega_{k,l}$; the integer multiples of the period coincide with the elements of the time invariant set. For short times $\tau$ such that $\sin \omega_{k,l} \tau \simeq \omega_{k,l} \tau$, the expectation value becomes
\begin{equation}
    \braket{\varphi_{k,l}|\operator{T}_{t(\tau)}| \varphi_{k,l}} = \tau \, ,
\end{equation}
in accordance with our results above. This would then hint that elements of the canonical domain such as $\ket{\varphi_{k,l}}$ can act as a clock state for a quantum clock. However, the time-energy uncertainty relation for small $\tau$ is not saturated. Now, we consider how we can saturate the uncertainty relation by projective means.

Given $k$ and $l$, consider the closed subspace spanned by the states $\ket{k}$ and $\ket{l}$. We denote this subspace as $\hilbertspace_{kl}$. The state $\ket{\varphi_{kl}} = 2^{-1/2} (\ket{k} - \ket{l})$ is the only element in $\hilbertspace_{kl}$ that also belongs to the canonical domain $\domain_c$. The set $\mathscr{T}_{kl}$ \eqref{eq:tkl} becomes the time invariant set in $\hilbertspace_{kl}$, and so for all $t$ in $\mathscr{T} = \mathscr{T}_{kl}$, $\operator{U}_t\ket{\varphi_{kl}}$ is also in $\domain_c$. Let $\operator{P}_{kl}$ be the projection operator onto $\hilbertspace_{kl}$. Then, let us consider the bounded operators $\operator{T}_{kl} = \operator{P}_{kl} \operator{T} \operator{P}_{kl} = i \omega_{kl}^{-1} (\ket{k}\bra{l} - \ket{l}\bra{k})$ and $\operator{H}_{kl} = \operator{P}_{kl} \operator{H} \operator{P}_{kl} = E_k \ket{k}\bra{k} + E_l \ket{l}\bra{l}$. Since $\operator{T}$ and $\operator{H}$ are canonically conjugate in $\domain_c$, we expect that $\operator{T}_{kl}$ and $\operator{H}_{kl}$ continue to satisfy the canonical commutation relation in the subspace of $\hilbertspace_{kl}$ spanned by $\ket{\varphi_{kl}}$, i.e., $(\operator{T}_{kl} \operator{H}_{kl} - \operator{H}_{kl} \operator{T}_{kl})\ket{\varphi_{kl}} = i\hbar \ket{\varphi_{kl}}$.

The significance of the projection can be seen from the fact that the expectation values of $\operator{T}$ and $\operator{T}_{kl}$ are equal, but with different variances, with $\operator{T}_{kl}$ having the smaller variance. This can be seen as follows. Note that $\operator{U}_t \ket{\varphi_{kl}} = \operator{P}_{kl} \operator{U}_t \operator{P}_{kl} \ket{\varphi_{kl}}$. Once again, let $t = \tau + 2n\pi/\omega_{kl}$ where $2n\pi/\omega_{kl} \in \mathscr{T} = \mathscr{T}_{kl}$ for integer $n$. For $\operator{T}_{kl}(t(\tau)) = \operator{U}_{t(\tau)}^\dagger \operator{T}_{kl} \operator{U}_{t(\tau)}$, we obtain the same expectation value as above
\begin{equation}
    \braket{\varphi_{kl}|\operator{T}_{kl}(t(\tau))|\varphi_{kl}} = \frac{1}{\omega_{kl}} \sin \omega_{kl} \tau \, .
\end{equation}
Now, the variance of $\operator{T}_{kl}$ is given by
\begin{equation}
    \Delta T_{kl}^2(t(\tau)) = \frac{1}{\omega_{kl}^2} \cos^2 \omega_{kl} \tau \, .
\end{equation}
With $\Delta H_{kl}(t(\tau)) = \sqrt{\braket{\varphi_{kl}|\operator{H}_{kl}^2(t(\tau))|\varphi_{kl}} - \braket{\varphi_{kl}|\operator{H}_{kl}(t(\tau))|\varphi_{kl}}^2} = \frac{1}{2} (E_k - E_l)$, we get the uncertainty relation
\begin{equation}
    \Delta T_{kl}(t(\tau)) \Delta H_{kl}(t(\tau)) = \frac{\hbar}{2} \abs{\cos \omega_{kl} \tau} \, .
\end{equation}
While the relation above is at its maximum of $\hbar/2$ when $\tau = 0$, i.e., when $t = 2n\pi/\omega_{kl} \in \mathscr{T}$ ($n = 0, \pm 1, \pm 2, \dotsc$), one may notice that it approaches zero as $t$ goes away from $\mathscr{T}$. This is because the state approaches the eigenkets of $\operator{T}_{kl}$, and thus, the uncertainty $\Delta T_{kl}$ goes to zero. While it may seem like a violation of the expected uncertainty relation, one only needs to recall that the operator $\operator{T}_{kl}$ is only a TECCR solution for times in the neighborhood of the time invariant set $\mathscr{T}$. Outside those times, the TECCR fails to hold, and the expectation value is not the parametric time. As with a similar discussion in \cite{garrison1970canonically}, when the state is not anymore in the canonical domain, we do not expect the usual form $\Delta T \Delta H = \hbar/2$ as our uncertainty relation, as the operator fails to be canonically conjugate to the Hamiltonian. 

We then only look at the case when $\operator{T}_{kl}$ is a TECCR solution, and thus, for a given $n$ and for small $\tau$, we have the desired results
\begin{equation}
    \braket{\varphi_{kl}|\operator{T}_{kl}(t(\tau))|\varphi_{kl}} = \tau \, ,
\end{equation}
\begin{equation}
    \Delta T_{kl}(t(\tau)) \Delta H_{kl}(t(\tau)) = \frac{\hbar}{2} \, .
\end{equation}
We see that the time-energy uncertainty relation is saturated in the neighborhood of $\mathscr{T}$ upon projecting the time operator and the Hamiltonian onto a closed subspace of $\hilbertspace$.

Therefore, the projection of the characteristic time operator on the Hilbert space spanned by $\ket{k}$ and $\ket{l}$ can serve as a quantum clock. The canonical domain of the TECCR solution would then dictate the clock states. In this case, consider the time-evolved states of the canonical domain $\ket{\Lambda(\tau)} = 2^{-1/2} ( \exp(-i\omega_k \tau) \ket{k} - \exp(-i\omega_l \tau) \ket{l})$. Firstly, it satisfies the property $\operator{U}_t \ket{\Lambda(\tau)} = \ket{\Lambda(\tau+t)}$. Second, local to each time in the time invariant set $\mathscr{T}=\mathscr{T}_{kl}$, the two-dimensional projection of the characteristic time operator gives the expectation value $\tau$, and the uncertainty principle is saturated. This then suggests that the states $\ket{\Lambda(\tau)}$ have the required properties of a clock state for times $\tau$ in the neighborhood of $\mathscr{T}$. In other words, the clock states are the states that are in the neighborhood of the canonical domain.

\section{Quantum clock} \label{sec:qclock}
Consider the Hamiltonian
\begin{equation}
    \operator{H}_{10} = \mqty(E_1 && 0 \\ 0 && E_0) \, ,
\end{equation}
on the Hilbert space $\hilbertspace_{10}$ spanned by the energy eigenkets $\ket{0} = \smqty(0 \\ 1)$ and $\ket{1} = \smqty(1 \\ 0)$, where $E_1 > E_0$ by assumption. The two-dimensional projection of the characteristic time operator onto $\hilbertspace_{10}$ is
\begin{equation}
    \operator{T}_{10} = \mqty(0 && \frac{i\hbar}{E_1 - E_0} \\ \frac{i\hbar}{E_0 - E_1} && 0) \, .
\end{equation}
Consider the state
\begin{equation}
    \ket{\varphi_{10}} = \frac{1}{\sqrt{2}} \mqty(1 \\ -1) \, ,
\end{equation}
which is an element of both $\hilbertspace_{10}$ and $\domain_c$. Then $(\operator{T}_{10}\operator{H}_{10} - \operator{H}_{10}\operator{T}_{10})\ket{\varphi_{10}} = i\hbar\ket{\varphi_{10}}$ is satisfied. We obtain the time invariant set
\begin{equation}
    \mathscr{T}_{10} = \qty{t: t = \frac{2n\pi\hbar}{E_1 - E_0}, n = 0, \pm 1, \pm 2, \dotsc} \, .
\end{equation}
Thus, with $t = \tau + 2n\pi\hbar/(E_1 - E_0)$ for integer $n$ and for small $\tau$ (i.e., when $\sin (\hbar \tau/(E_1 - E_0)) \simeq \hbar \tau/(E_1 - E_0)$ and $\cos (\hbar \tau/(E_1 - E_0)) \simeq 1$), we have
\begin{equation}
    \braket{\varphi_{10}|\operator{T}_{10}(t(\tau))|\varphi_{10}} = \tau \, ,
\end{equation}
\begin{equation}
    \Delta T_{10}(t(\tau)) \Delta E_{10} (t(\tau)) = \frac{\hbar}{2} \, ,
\end{equation}
i.e., in the neighborhood of the time invariant set, the expectation value gives the parametric time $\tau$, and the uncertainty principle is saturated.

We shall now show the measurement scheme that will reveal the physical interpretation of the characteristic time operator $\operator{T}$. Consider a physical observable $\operator{A}$ that commutes with $\operator{T}$, i.e., $[\operator{A},\operator{T}]=0$, and thus, both have a common set of eigenfunctions. Measurement of this other observable $\operator{A}$ then provides the link in measuring $\operator{T}$, and thus, determining the parametric time $\tau$. This is analogous to how one can measure time using a traditional clock: measurement of the position of the clock's hands will determine the time. One can then think of $\operator{A}$ as the \textit{clock-hand operator} or the clock pointer.

One such \textit{clock-hand} observable can be noticed when we rewrite the characteristic time as
\begin{equation}
    \operator{T}_{10} = \frac{\hbar}{E_0 - E_1} \sigma_y \, ,
\end{equation}
where $\sigma_y = \smqty(\pmat{2})$ is the Pauli matrix in $y$. We then look at one specific system as an example: the Larmor clock. Let $\operator{S}_y = (\hbar/2) \sigma_y$ be the observable corresponding to the spin in the $y$-direction. Consider the Hamiltonian $\operator{H} = \omega \operator{S}_z$, where $\operator{S}_z = (\hbar/2) \sigma_z = \frac{\hbar}{2} \smqty(\pmat{3})$. This then models the familiar Larmor precession, wherein a stationary spin-$1/2$ particle is under a static uniform magnetic field in the $+z$-direction with Larmor frequency $\omega$. We then write our characteristic time operator as $\operator{T} = -(2/\hbar\omega) \operator{S}_y$. Let the state initially be spin-down in the $x$-direction, i.e., $\ket{\operator{S}_{x,-}} = \frac{1}{\sqrt{2}} \smqty(1 \\ -1) $, and let it evolve through time, $\ket{\operator{S}_{x,-}(t)}=\operator{U}_t \ket{\operator{S}_{x,-}}$. Note that $\ket{\operator{S}_{x,-}}$ is inside the canonical domain $\domain_c$ and will serve as our clock state; as it evolves to $\ket{\operator{S}_{x,-}(t)}$, it leaves the canonical domain, returning only at times in the time invariant set. We then measure $\operator{S}_y$, which will act as the \textit{hand of our clock}, measurement of which helps us access the parametric time. Let $t = \tau + 2n\pi/\omega$ ($n = 0, 1, 2, \dotsc$), where $2n\pi/\omega$ are elements of the time invariant set, which are just integer multiples of the Larmor period. We repeat this procedure multiple times to obtain $\braket{\operator{S}_{x,-}(t(\tau))|\operator{S}_y|\operator{S}_{x,-}(t(\tau))}$. With the relationship between $\operator{T}$ and $\operator{S}_y$, we now know that for small $\tau$, we get the parametric time
\begin{equation} \label{eq:timespin}
    \tau = -\frac{2}{\hbar\omega} \braket{\operator{S}_{x,-}(t(\tau))|\operator{S}_y|\operator{S}_{x,-}(t(\tau))} \, ,
\end{equation}
and that the uncertainty relation is simultaneously saturated
\begin{equation}
    \Delta T(t(\tau)) \Delta H(t(\tau)) = \frac{\hbar}{2} \, .
\end{equation}

Therefore, for a Larmor clock with period $2\pi/\omega$, the \textit{ticks} of the clock are when the state is at the clock state $\ket{\operator{S}_{x,-}}$. This state $\ket{\operator{S}_{x,-}}$ is inside the canonical domain $\domain_c$, it goes outside $\domain_c$ as time goes on, and then returns back to the canonical domain to become $\ket{\operator{S}_{x,-}}$ once again. The time it takes for it to return is given by the system's period or the recurrence time, which is equal to this clock's time resolution. One can access the parametric time $t$ by measuring time $\tau$ in the neighborhood of each \textit{tick} of the clock via \eqref{eq:timespin}, which is related to $t$ by $\tau = t \mod(2\pi/\omega)$. If one can count the number of cycles $n$, we obtain $2n\pi/\omega$, and get $t = \tau + 2n\pi/\omega$. 

One may recognize \eqref{eq:timespin} to have a very similar form to the mean Larmor precession transmission and reflection times for particles incident on a barrier with a magnetic field inside the barrier \cite{mayato2008quantum}. As another point of comparison, similar to how our time operator $\operator{T}$ is proportional to the compatible observable $\operator{S}_y$, the harmonic oscillator quantum clock in \cite{garrison1970canonically} also has the self-adjoint time operator of the form $\operator{T} = (\hbar\omega)^{-1} \operator{F}$ to tell the clock time, where $\operator{F}$ is the phase operator, with the Hamiltonian $\operator{H}=\hbar\omega\operator{N}$ where $\operator{N}$ is the number operator. Finally, we compare with the results of \cite{caballar2009characterizing} and look at the dynamics of the eigenstates of $\operator{T}$, $\ket{\operator{S}_{y,\pm}}$, which will serve as the clock states. One difference though is that the (infinite-dimensional) characteristic time operator was studied in \cite{caballar2009characterizing}, while in this section, we only looked at its two-dimensional projection. From the previous section, we know that the uncertainty in time does go to a minimum at specific times (i.e., for times in the time invariant set); in fact, for the two-dimensional projection of $\operator{T}$, the uncertainty goes to zero at the eigenstates of $\operator{T}$. We observe that the time it takes to evolve from the eigenstate $\ket{\operator{S}_{y,+}}$ to the other eigenstate $\ket{\operator{S}_{y,-}}$ is $\pi/\omega$, which is not equal to the difference in the eigenvalues of $\operator{T}$, $2/\omega$, though this may be due to taking the two-dimensional projection of $\operator{T}$, and thus the eigenvalue difference is not sufficiently small.

\section{Discussion} \label{sec:disc}
In \cite{galapon2006could}, a TECCR solution is originally characterized by the triple $\mathcal{C}(\operator{T},\operator{H},\domain_c)$. The results in this work suggest that a solution to the TECCR should instead be characterized as $\mathcal{C}(\operator{T},\operator{H},\domain_c,\mathscr{T})$, since $\operator{T}$ and $\operator{H}$ form a canonical relation in $\domain_c$ only in the neighborhood of times in $\mathscr{T}$. This emphasizes the fact that there exists TECCR solutions that are not covariant, wherein $\operator{T}$ acts as a legitimate time operator only in the neighborhood of $\mathscr{T}$. Even though covariance is violated, we have shown that self-adjoint time operators are still physically useful! Therefore, covariance is not a necessary property that should be forced on all time operators.

We see that the two-dimensional projection of the characteristic time operator behaves as a quantum clock. The clock states are elements of the canonical domain $\domain_c$, as well as the states very close to $\domain_c$. As the states evolve, they move away from $\domain_c$, and returns to $\domain_c$ only for times in the time invariant set $\mathscr{T}$. In the neighborhood of the times in $\mathscr{T}$ (i.e., when the state is near the canonical domain $\domain_c$), the time operator moves in step with parametric time, its expectation value gives the parametric time, and the time-energy uncertainty relation is saturated. As we can see, the constructed clock does not work for all times. Since the desired quantum clock properties only hold in the neighborhood of a set of measure zero, we interpret it as a limit imposed by quantum mechanics in measuring the time using the system's characteristic time. This can be thought of as a phenomenon of the quantization of time, in the sense that we can only measure time in discrete steps; in our case, we can actually measure time in the neighborhood of those discrete steps. We are then faced with the problem of nonlinearity when measuring time far away from these steps.

The clock arising from the characteristic time operator can be thought of as the system's \textit{characteristic} quantum clock. Once the Hamiltonian is determined, one automatically obtains the characteristic time operator (which will be measured to get the clock time), the canonical domain (which will determine the clock states), and the time invariant set (which will determine the clock's time resolution). For example consider again the clock state $2^{-1/2} (\ket{k} - \ket{l})$ in $\domain_c$. The characteristic quantum clock has time resolution $\delta t = 2\pi\hbar/(E_k - E_l)$ arising from the time invariant set; in truth, the clock does not only tell the time at integer multiples of the resolution, but also at its neighborhood. There is also an uncertainty in the measurement of time $\Delta T = \hbar / (E_k - E_l)$ in the neighborhood of the time resolution via the uncertainty principle. For a semibounded Hamiltonian, we can have an arbitrary energy difference, implying a two-fold effect: the ability to construct a clock with arbitrary resolution $\delta t$ and the ability to measure time in the neighborhood of the resolution with arbitrary accuracy $\Delta T$.

In the literature, construction of a quantum clock did not need a self-adjoint time operator. One simply needs a system that evolves from one state (say, the ground state $\ket{0}$) to another (say, the excited state $\ket{1}$), and find a dynamical observable with a simple dependence on parametric time or clock time. One can then measure this dynamical observable to infer the clock time. While not a strict requirement, clocks are usually periodic systems, e.g., angle eigenkets $\ket{\theta}$ where $\theta$ runs through $[0,2\pi]$ with periodic boundary conditions, spin precession, harmonic oscillators, and so on. Clocks (especially classical clocks) can then be constructed using a system's natural frequency, and thus it is characterized by the period of the system or by the recurrence time---the time it takes for a state to return to itself. Suppose the clock-hand observable depends on time $t$ linearly. One can then measure this observable to obtain $\tau = t \mod T$ where $T$ is the period of the system. Suppose the experimenter knows when the clock was turned on, and suppose they can also count how many cycles $n$ happened. One can then obtain the time that has elapsed $\tau + nT$. 

In measuring observables in quantum mechanics, either (i) one appeals to the direct quantum representation of the observable, or (ii) one appeals to another observable with a known relationship with the former. In the measurement of observables via the second kind, either (a) there exists a self-adjoint representation of the observable and another operator which commutes with it, or (b) it may be otherwise (which could be the case when there is no self-adjoint representation for the observable). Historically, it was believed that no self-adjoint time operator exists, thus, quantum clock measurements can only be of the latter type. In this work, we show that for semibounded discrete Hamiltonians with some growth condition, there actually always exists an underlying self-adjoint time operator providing a characteristic quantum clock. This self-adjoint characteristic time operator then corresponds to the ``clock time" observable, and we can find a commuting operator (such as spin in the Larmor clock example) which we can measure and thus know the parametric time via its relationship with the time operator. This parametric time can be accessed at its period, as the operator is canonically conjugate to the Hamiltonian in the set of times which are integer multiples of the period. Not only that, the operator can also access time outside the ticks (though only barely), as one can also measure the time in the neighborhood of the period. For now, this may seem to be the best that a quantum characteristic time can give us, denying us access of the parametric time when measuring far from its period. Quantum mechanics, while allowing clock measurements via non-covariant time operators, still limits our ability in measuring time.


\end{document}